\documentstyle[aps,prl,preprint]{revtex}

\begin{document}

\preprint{\vbox{\baselineskip=3ex
                \hbox{KEK-TH-478}
                \hbox{KEK preprint 96-5}
                \hbox{April 1996}
                \hbox{H}}
                \bigskip}
\draft
\title{
$\mu\rightarrow e\gamma$ Search with Polarized Muons}
\author{Yoshitaka Kuno and Yasuhiro Okada}
\address{Department of Physics, National Laboratory for High Energy
Physics (KEK),\\ Tsukuba, Ibaraki, Japan 305}

\maketitle

\begin{abstract}

A search for the lepton-flavor violating $\mu^{+}\rightarrow
e^{+}\gamma$ decay using polarized muons is proposed. By measuring the
angular distribution of $e^{+}$s with respect to the muon spin
direction, in particular antiparallel $e^{+}$s, the serious physics
background from $\mu^{+}\rightarrow e^{+}\nu\overline{\nu}\gamma$ as
well as accidental background from normal muon decay accompanied by a
high-energy photon can be suppressed significantly. In addition to the
enhancement of the sensitivity, the angular distribution would
discriminate among different extensions to the Standard Model, once
the signal is observed.

\end{abstract}

\bigskip



Lepton flavor violation (LFV) has been attracting much attention
recently, since several extensions to the Standard Model predict large
LFV rates. Among the various theoretical models with LFV, models of
supersymmetric grand unification (GUT) in particular predict a
branching ratio only one or two order magnitudes lower than the
current experimental limits \cite{hall86,barb95}. A search for LFV in
low energy processes would have the potential to test unification at
very high energy. Compared to other LFV processes of interest such as
$\mu^{+}\rightarrow 3e$, $\mu^{-} - e^{-}$ conversion in a nucleus,
$\tau\rightarrow \mu\gamma$, etc. $\mu^{+}\rightarrow e^{+}\gamma$ is
known to have a higher sensitivity to supersymmetric unification
\cite{barb95}.

The previous experiments searching for $\mu^{+}\rightarrow
e^{+}\gamma$ have been done with a surface muon beam of low-energy (4
MeV) positive muons from pions stopped in the pion production
target. Due to their production mechanism, surface muons are
originally 100 \% polarized, antiparallel to their flight
direction. However, none of the past experiments used the
polarization, but rather depolarized the muons by using, for instance,
a polystyrene muon-stopping target \cite{bolt88}. Here, the importance
of the muon polarization to search for $\mu^{+}\rightarrow
e^{+}\gamma$ will be presented. For instance, the measurement of the
angular distribution of $\mu^{+}\rightarrow e^{+} \gamma$ with respect
to the muon-polarization direction would give information to
discriminate among various models, since different theoretical models
predict a different helicity of $e^{+}$ in $\mu^{+}
\rightarrow e^{+} \gamma$, as shown below.

Supersymmetric unified theories predict large rates for LFV processes
such as $\mu^{+}\rightarrow e^{+}\gamma$ as a consequence of the large
top-quark Yukawa coupling. The SU(5) supersymmetric (SUSY) GUT model
predicts a branching ratio between $10^{-15}$ to $10^{-13}$ for the
singlet selectron mass of $m_{\tilde{e}_{R}}$ of 100 to 300 GeV, and
the SO(10) SUSY-GUT models give an even larger value of $10^{-13}$ to
$10^{-11}$ \cite{barb95}.  In these supersymmetric unified models,
interactions at the GUT energy scale could induce lepton flavor mixing
in the left-handed and/or right-handed slepton sectors. As a
consequence, LFV processes such as $\mu \rightarrow e \gamma$ occur at
a low energy, and they therefore carry information on the interactions
at the GUT energy scale. The helicity of $e^{+}$ in $\mu^{+}
\rightarrow e^{+} \gamma$ is subject to the mechanism of flavor
mixing in the slepton sectors. For instance, the minimal SU(5)
SUSY-GUT model introduces a lepton flavor mixing only on the
right-handed slepton sector, $\tilde{e}_{R}$, and therefore, only
$\mu^{+} \rightarrow e^{+}_{L}\gamma$ occurs, where $e^{+}_{L}$ is a
left-handed positron. On the other hand, the SO(10) SUSY-GUT models
have a lepton flavor mixing on the $\tilde{e}_{L}$ sector as well as
the $\tilde{e}_{R}$ sector, giving rise to $\mu^{+}\rightarrow
e^{+}_{R}\gamma $ as well as $\mu^{+} \rightarrow
e^{+}_{L}\gamma$. Furthermore, some non-unified supersymmetric
extensions to the Standard Model with heavy right-handed neutrinos
also expect a large branching ratio of orders of $10^{-13}$ to
$10^{-11}$\cite{hisa95} and $\mu^{+}\rightarrow e^{+}_{R} \gamma$ is
predicted. Some non-supersymmetric extensions to the Standard Model,
such as left-right symmetric models and extra Higgs models, also
predict a sizeable branching ratio of $\mu^{+} \rightarrow
e^{+}\gamma$ \cite{verg86}.

However, to improve sensitivity and measure angular distribution,
major background decays to a search for $\mu^{+}\rightarrow
e^{+}\gamma$, in particular their angular distribution with respect to
the muon-polarization direction, have to be investigated. Those
backgrounds are a physics background from the radiative muon decay
$\mu^{+}\rightarrow e^{+}\nu\overline{\nu}\gamma$, and an accidental
background of the normal muon decay $\mu^{+}\rightarrow
e^{+}\nu\overline{\nu}$ accompanied by a high energy photon. In this
letter, we propose the use of polarized muons to search for
$\mu^{+}\rightarrow e^{+}\gamma$ to improve sensitivity against those
backgrounds, and especially emphasize the importance of the selective
measurement of $e^{+}$s going opposite to the muon-polarization
direction, which would reduce the backgrounds significantly and allow
the measurement with a higher sensitivity in future experiments.

In general, the Lagrangian for $\mu^{+}\rightarrow e^{+}\gamma$ decay
can be given \cite{wein59} with an explicit helicity expression by

\begin{equation}
{\cal L}=
A_{R}~\overline{\mu}_{R}\sigma^{\mu\nu}e_{L}F_{\mu\nu} +
A_{L}~\overline{\mu}_{L}\sigma^{\mu\nu}e_{R}F_{\mu\nu} + h.c.
\label{eq:lag}
\end{equation}

\noindent 
where $\mu_{L}(\mu_{R})$ is a left-handed (right-handed) muon and
$e_{L}(e_{R})$ is a left-handed (right-handed) electron. $A_{R}$ and
$A_{L}$ are coupling constants. $F_{\mu\nu}$ is
$\partial_{\mu}A_{\nu}-\partial_{\nu}A_{\mu}$ and $A_{\mu}$ is an
electromagnetic field. For polarized muons, from Eq.(\ref{eq:lag}),
the angular distribution of $e^{+}$ in $\mu^{+}\rightarrow
e^{+}\gamma$ with respect to the direction of $\mu^{+}$ spin direction
can be given by

\begin{equation}
d\Gamma(\mu^{+}\rightarrow e^{+}\gamma) = {1\over 8\pi}\bigl({1 -
{m_{e}^2\over m_{\mu}^{2}}}\bigr)^{3} m_{\mu}^{3} d(\cos\theta)
\times
\Bigl[ 
|A_R|^2(1-P_{\mu}\cos\theta) + |A_L|^2(1+P_{\mu}\cos\theta) 
\Bigr]
\end{equation}

\noindent where $\theta$ is the angle between the muon spin and the
$e^{+}$ direction. $P_{\mu}$ is the muon spin polarization, which for
surface muons is almost 100 \%. $m_{\mu}$ and $m_{e}$ are the muon
mass and the positron mass, respectively. The minimal SU(5) SUSY-GUT
model predicts a non-zero $A_L$ and a vanishing $A_R$, yielding a
($1+P_{\mu}\cos\theta$) distribution. On the other hand, the simplest
version of the SO(10) SUSY-GUT models predict approximately-equal
helicity amplitudes for right-handed and left-handed $e^{+}$s
($A_L\approx A_R$) \cite{barb95}, resulting in an almost uniform
angular distribution. For some non-unified supersymmetric
models\cite{hisa95}, $A_L$ is vanishing but $A_R$ is non-zero, giving
a ($1-P_{\mu}\cos\theta$) distribution. Therefore, the measurement of
the angular distribution of $e^{+}$ with respect to the direction of
muon polarization would provide a means to discriminate among these
models clearly when the signals are observed.

The event signature of $\mu^{+}\rightarrow e^{+}\gamma$ is that the
energies of both $e^{+}$ and photon are equal to a half of the muon
mass ($m_{\mu}/2 =$ 52.8 MeV) and they are collinear back to back. The
current experimental upper limit is $4.9 \times 10^{-11}$ at 90 \%
confidence level\cite{bolt88}.

The major physics backgrounds to the search for $\mu^{+}
\rightarrow e^{+} \gamma$ decay is radiative muon decay, $\mu^{+}
\rightarrow e^{+}\nu\overline{\nu}\gamma$ (branching ratio = 1.4 \%
for $E_{\gamma} >$ 10 MeV). When the $e^{+}$ and photon are emitted
back-to-back with two neutrinos carrying off little energy, it becomes
a serious physics background to $\mu^{+}\rightarrow e^{+} \gamma$. The
differential decay width of this radiative decay was calculated as a
function of $e^{+}$ energy ($E_{e}$) and photon energy ($E_{\gamma}$)
normalized to their maximum energies (of $m_{\mu}/2$), namely $x =
2E_{e}/m_{\mu}$ and $y = 2E_{\gamma}/m_{\mu}$, where $x$ and $y$ range
from 0 to 1 \cite{fron59,ecks59}. As a background to
$\mu^{+}\rightarrow e^{+}\gamma$, the kinematic case when $x\approx1$
and $y\approx1$ is important. Its angular distribution has to be
examined especially. In an approximation of the limit of $x\approx1$
and $y\approx1$ with an angle between $e^{+}$ and photon
($\theta_{e\gamma})$ of almost $180^{\circ}$, the differential decay
width of $\mu^{+}\rightarrow e^{+}\nu\overline{\nu}\gamma$ decay is
given by

\begin{eqnarray}
&&d\Gamma(\mu\rightarrow e\nu\overline{\nu}\gamma)
\cong {G_{F}^2 m_{\mu}^5 \alpha \over 3\times 2^{8}\pi^{4}} \times \cr
&&\Bigl[ (1-x)^{2}(1-P_{\mu}\cos\theta) + 
\Bigl( 4(1-x)(1-y) - {1\over2}z^{2} \Bigr)(1+P_{\mu}\cos\theta)
\Bigr]dxdyzdzd(\cos\theta)
\label{eq:dif}
\end{eqnarray}

\noindent 
where $G_{F}$ is the Fermi coupling constant, $\alpha$ is the
fine-structure constant, $z = \pi-\theta_{e\gamma}$, and $\cos z$ is
expanded in a polynomial of $z$ since $z$ is small. In
Eq.(\ref{eq:dif}), only the terms of up to the second order of a
combination of $(1-x)$, $(1-y)$ and $z$ are kept. At $x\approx1$ and
$y\approx 1$, the effect of the positron mass is found to be very
small, of order $(m_{e}/m_{\mu})^2$, and therefore neglected in
Eq.(\ref{eq:dif}). The first term in Eq.(\ref{eq:dif}) represents the
$e^{+}$ being emitted preferentially opposite to the muon spin
direction, whereas in the second term the $e^{+}$ is emitted along the
muon-spin direction. When $x=1$ and $y=1$ exactly, this differential
decay width vanishes. But in a real experiment, finite detector
resolutions introduce background events which would ultimately limit
the sensitivity of a search for $\mu^{+}\rightarrow e^{+}\gamma$.

Given the detector resolutions, the sensitivity limitation from
$\mu^{+}\rightarrow e^{+}\nu\overline{\nu}\gamma$ background decay can
be evaluated by integrating the differential decay width in
Eq.(\ref{eq:dif}) over the resolutions, or more precisely, over the
kinematic box region of the signal which is determined by the detector
resolutions \cite{dete96}. Take $\delta x$, $\delta y$ and $\delta z$
to be the kinematic range of the signal region for $e^{+}$ energy
($1-\delta x \leq x \leq 1$), that for photon energy ($1-\delta y\leq
y \leq 1$) and that for the angle of $z = \pi-\theta_{e\gamma}$ ($0\leq
z \leq \delta z$), respectively. The integration was done with
consideration of the kinematics constraints among $x$, $y$ and
$z$. Namely, if $\delta x$ and $\delta y$ are small, the allowed range
of $z$ is determined to be $0\leq z \leq 2\sqrt{(1-x)(1-y)}$, instead
of $\delta z$.  The partial branching ratio after the integration is
given by

\begin{eqnarray}
dB(\mu\rightarrow e\nu\overline{\nu}\gamma) &=&
{1 \over \Gamma(\mu\rightarrow e\nu\overline{\nu})} 
\int^{1}_{1-\delta x} dx \int^{1}_{1-\delta y} dy
\int^{\min(\delta z, 2\sqrt{(1-x)(1-y)})}_{0} dz
~~{d\Gamma(\mu\rightarrow e\nu\overline{\nu}\gamma) \over dxdydz} \cr &=&
{\alpha\over16\pi}
\Bigl[ 
J_{1}(1-P_{\mu}\cos\theta) 
+ J_{2}(1 + P_{\mu}\cos\theta) 
\Bigr]d(\cos\theta),
\label{eq:par}
\end{eqnarray}

\noindent where $\Gamma(\mu\rightarrow e\nu\overline{\nu})$ is the total
muon decay width. For the case of $\delta z >2\sqrt{\delta x\delta
y}$, $J_1$ and $J_2$ are given by

\begin{equation}
J_1 = (\delta x)^4(\delta y)^2 \quad {\rm and} 
\quad J_2 = {8\over3}(\delta x)^3(\delta y)^3,
\label{eq:par1}
\end{equation}

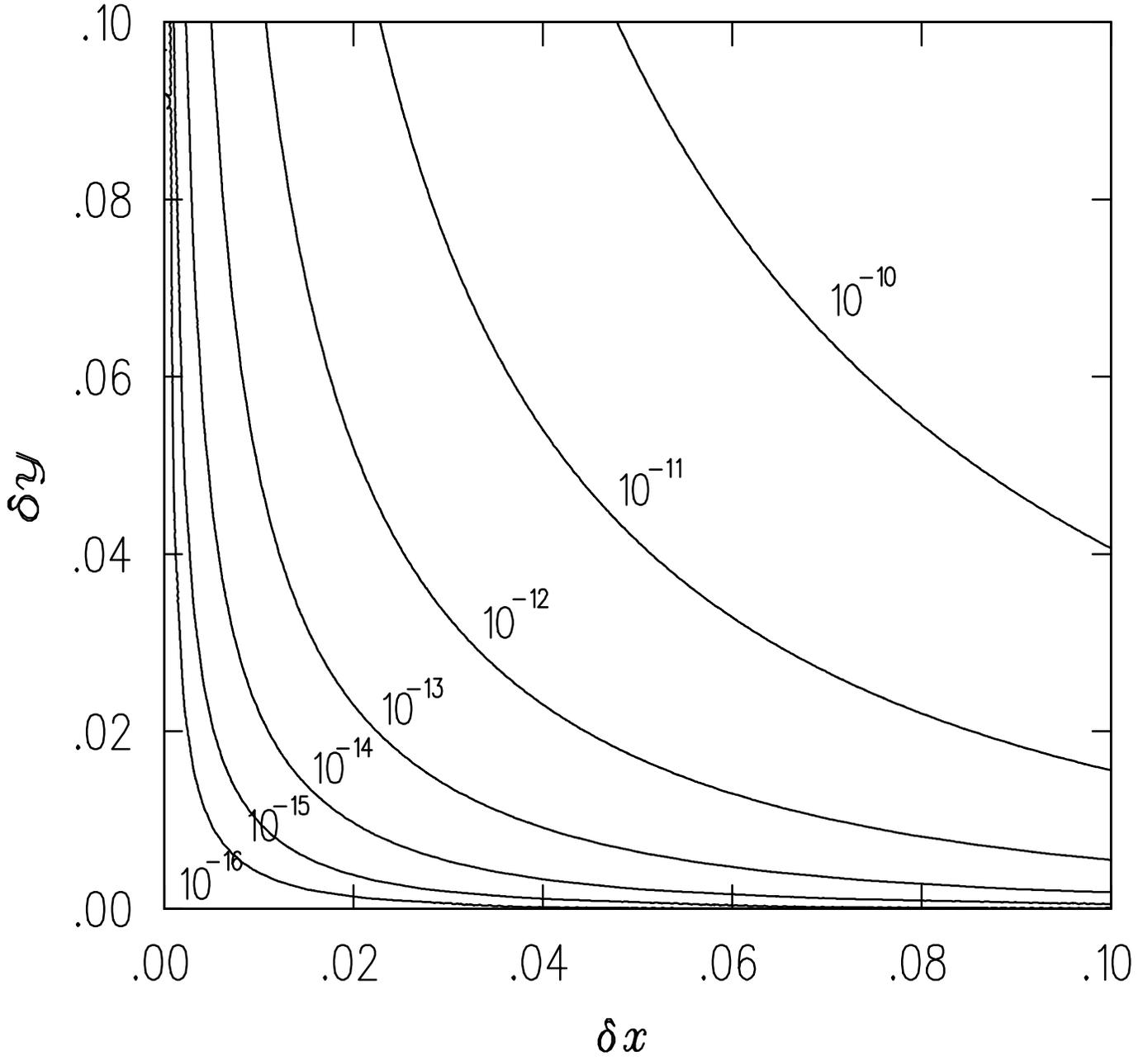
\begin{figure}
\caption{Sensitivity limitation of the branching ratio of
$\mu^{+}\rightarrow e^{+}\gamma$ imposed by $\mu^{+}\rightarrow
e^{+}\nu\overline{\nu}\gamma$ decay for the case of unpolarized muons 
as a function of $\delta x$ and $\delta y$.}
\label{fg:sen}
\end{figure}

\noindent 
{\em i.e.}, they can be reduced as the sixth power of a combination of
$\delta x$ and $\delta y$. In Fig.\ref{fg:sen}, the sensitivity limit
imposed by the $\mu^{+}\rightarrow e^{+}\nu\overline{\nu}\gamma$ decay
with unpolarized muons is shown. From Fig.\ref{fg:sen}, it can be seen
that to achieve an sensitivity limit of a level of $10^{-15}$, both
$\delta x$ and $\delta y$ of an order of 0.01 are needed.

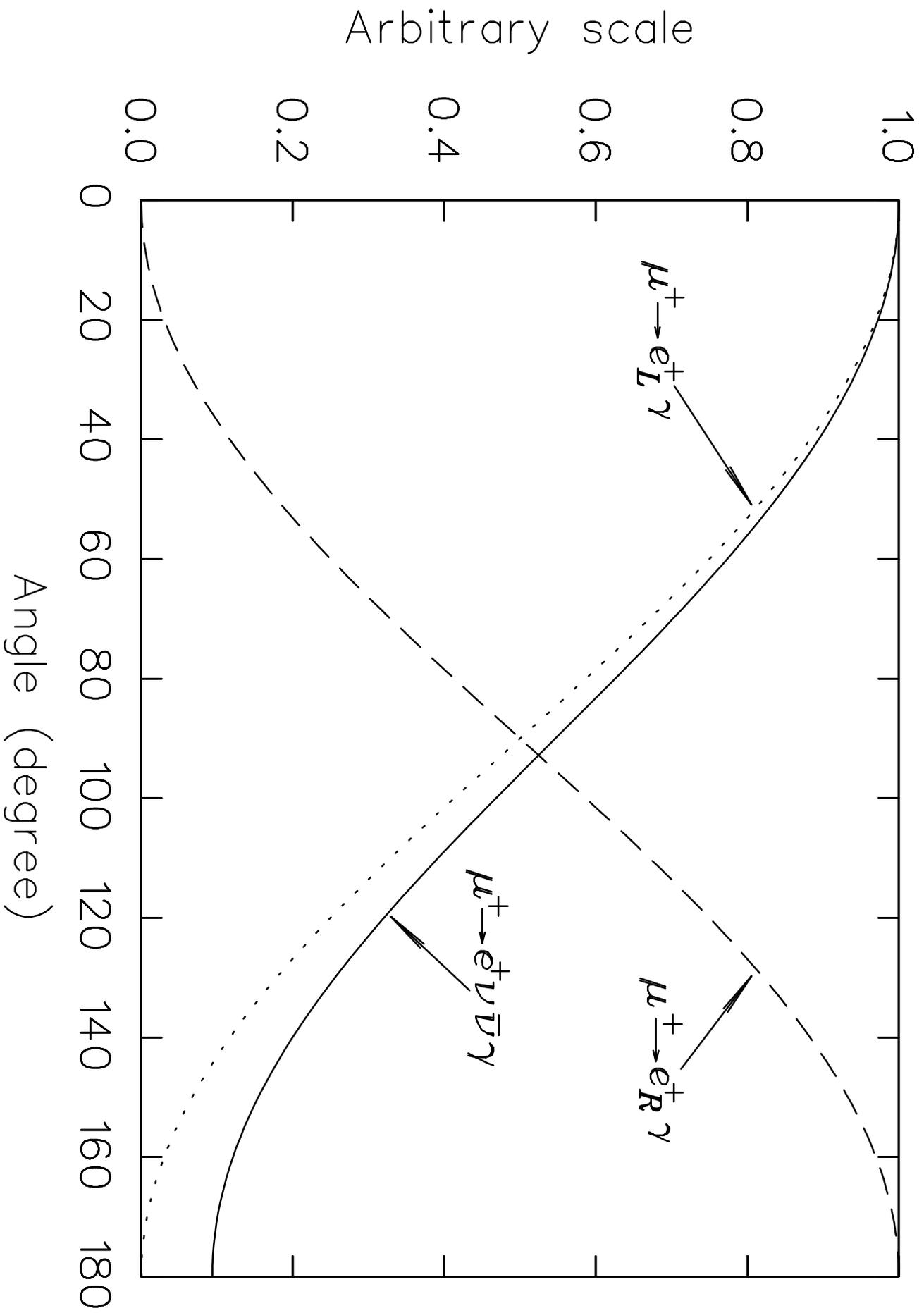
\begin{figure}
\caption{Angular distribution of $e^{+}$ in $\mu^{+}\rightarrow
e^{+}\nu\overline{\nu}\gamma$ decay with $\delta y/\delta x = 4$ as a
function of angle from the muon-polarization direction in a
solid line. A dot line and dash line show that of 
$\mu^{+}\rightarrow e^{+}_{L}\gamma$ and 
$\mu^{+}\rightarrow e^{+}_{R}\gamma$ decay,
respectively. A vertical scale is arbitrary.}
\label{fg:dis}
\end{figure}

Experimentally, improvement of the photon-energy resolution could be
more difficult than that of the $e^{+}$ energy resolution. For
instance, in the MEGA experiment which aims to measure the
branching ratio sensitive to a level of $6 \times 10^{-13}$
\cite{mega95}, the proposed numbers of the detector resolutions are
$\triangle E_{e} = 0.4$ MeV (FWHM) but $\triangle E_{\gamma} = 1.4$
MeV (FWHM). It is likely that the improvement of photon energy
resolution is limited. Assuming that $\delta y$ is worse than $\delta
x$ by a factor of several, together with the factor of 8/3, $J_2$ in
Eq.(\ref{eq:par}) could be an order magnitude larger than
$J_1$. Therefore, the distribution follows mostly
(1+$P_{\mu}\cos\theta$) as long as $\delta y > \delta
x$. Fig.\ref{fg:dis} shows the angular distribution of
$\mu^{+}\rightarrow e^{+}\nu\overline{\nu}\gamma$ with for instance
$\delta y/\delta x = 4$. It implies that if we measure selectively
$e^{+}$s in $\mu^{+}\rightarrow e^{+}\gamma$ going opposite to the
muon-polarization direction, the background from $\mu^{+}\rightarrow
e^{+}\nu\overline{\nu}\gamma$ will be reduced significantly.
Furthermore, by varying $\delta x$ and $\delta y$, the angular
distribution of the $\mu^{+}\rightarrow e^{+}\nu\overline{\nu}\gamma$
background can be changed according to Eq.(\ref{eq:par}), thus
providing another means to discriminate the signal from the
backgrounds.

When the angular resolution of $\delta z$ is better than the
kinematically-allowed angle of $2\sqrt{\delta x \delta y}$, the
distribution is given in Eq.(\ref{eq:par}) by

\begin{eqnarray}
J_1 &=& 
{8\over3}(\delta x)^3(\delta y)({\delta z\over 2})^2 
- 2(\delta x)^2({\delta z\over 2})^4 
+ {1\over 3}{1 \over (\delta y)^2}({\delta z\over 2})^8  \cr
J_2 &=& 
8(\delta x)^2(\delta y)^2({\delta z\over 2})^2
- 8(\delta x)(\delta y)({\delta z\over 2})^4 
+ {8\over 3}({\delta z\over 2})^6
\label{eq:par2}
\end{eqnarray}

\noindent Similarly to Eq.(\ref{eq:par1}), 
$J_2$ is larger than $J_1$ by roughly a factor of $\delta y/\delta
x$. The above argument is also valid in this case.

Another serious background to $\mu^{+}\rightarrow e^{+}\gamma$ is an
accidental coincidence of a high-energy $e^{+}$ in the normal muon
decay, $\mu^{+}\rightarrow e^{+}\nu\overline{\nu}$, with a high-energy
photon, likely from $\mu^{+}\rightarrow
e^{+}\nu\overline{\nu}\gamma$ \cite{bolt88,mega95}. This accidental
background can be reduced in principle by an improved timing
resolution of the experimental apparatus. However, at a high event
rate, it is desirable to reduce this background by some additional
means, and the use of polarized muons turns out to be very
effective. The high-energy $e^{+}$ in the normal muon decay ($x\approx
1$) is known to be emitted preferentially along the muon spin
direction, following an angular distribution of
$(1+P_{\mu}\cos\theta)$. When $e^{+}$s going opposite to the muon-spin
direction are selected, this background is also reduced.

Thus, all the serious backgrounds to $\mu^{+} \rightarrow e^{+}\gamma$
decay are expected to follow a $(1 + P_{\mu}\cos\theta)$ angular
distribution of for $e^{+}$s. The detection of $e^{+}$s emitted
opposite to the muon-spin direction would eliminate backgrounds,
whereas the acceptance of the signals for $\mu^{+}\rightarrow
e^{+}_{R}\gamma$ decay, which follows a $(1-P_{\mu}\cos\theta)$
distribution, is kept high. However, the sensitivity to
$\mu^{+}\rightarrow e^{+}_{L}\gamma$ will be the same as the case with
unpolarized muons. The angular distributions of $\mu^{+}\rightarrow
e^{+}_{R} \gamma$ and $\mu^{+}\rightarrow e^{+}_{L}\gamma$ are also
shown in Fig.\ref{fg:dis}. In the design of future experiments, the
solid-angle acceptance must be optimized with consideration of the
required background rejection and the available muon beam
intensity. For instance, taking a solid angle coverage of the
detection of $45^{\circ}$ ($60^{\circ}$) in polar angle with respect
to the direction antiparallel to the muon polarization would give an
improvement in a signal-to-background ratio of about 12 (about 7),
with a reasonable acceptance. Given a future increase in muon-beam
intensity, a next-generation experiment measuring the branching ratio
sensitive to a level of $10^{-14} - 10^{-15}$ without any backgrounds
could be possible with polarized muons.

Since a surface muon beam is already polarized to 100\%, the
measurement with polarized muons will require a suitable material as a
muon-stopping target which preserves the muon spin.  Further, since
the direction of muon polarization is antiparallel to its flight
direction, the installation of a muon-spin rotator \cite{mars92} which
rotates the muon-spin direction perpendicular to the beam direction
can be considered for more convenient detector arrangement.

In conclusion, a search for $\mu^{+}\rightarrow e^{+}\gamma$ decay
with polarized muons has the potential for improved sensitivity. The
selective measurement of $e^{+}$s going antiparallel to the muon-spin
direction would reduce the physics backgrounds from
$\mu^{+}\rightarrow e^{+}\nu\overline{\nu}\gamma$ decay as well as
from the accidental background of the normal muon decay accompanied by
a high-energy photon.  In addition, when the signal is observed, its
angular distribution would give a clear discrimination of models and a
significant test of supersymmetric unification.

\acknowledgments
The authors are grateful to Dr. J.A. Macdonald for the information on
a muon-spin rotator.  This work was supported in part by the
Grant-in-Aid of the Ministry of Education, Science, Sports and
Culture, Government of Japan.

\begin{references}
\bibitem{dete96}The size of the signal region is generally assigned
by a few times r.m.s of the detector resolution in order 
to keep a high acceptance of the signal.
\end{references}
\end{document}